%% file: ms.tex
 \documentclass[preprint2]{aastex}

\def\archivetally{1,353,228}
\def\catalogtally{933,818}

\def\robittally{18,949}
\def\robitmipstally{16,469}
\def\robitarchtally{16,079}
\def\robitcattally{14,926}
\def\hinzmipstally{120,883}
\def\hinzarchtally{82,832}
\def\hinzcattally{68,608}
\def\hinzcuttally{63,129}
\def\hinzcutarch{51,937}
\def\hinzcutmatch{39,910} 
\def\wisematch{368,956}

\def\matchtolerance{2$^{\prime\prime}$}
\def\arcsec{$^{\prime\prime}$}
\def\robitarchoffset{-0.07~mag}
\def\robitcatoffset{-0.09~mag}
\def\robitarchrms{0.19~mag}
\def\robitcatrms{0.14~mag}
\def\wisemagoffset{-0.07~mag}

\slugcomment{draft version \today}

\shorttitle{MIPSGAL PSC delivery}
\shortauthors{Gutermuth \& Heyer}

\begin{document}


\title{A 24\micron\ Point Source Catalog of the Galactic Plane from Spitzer/MIPSGAL.}


\author{Robert A. Gutermuth \and Mark Heyer}
\affil{Department of Astronomy, University of Massachusetts,
    Amherst, MA 01003}


\begin{abstract} 

In this contribution, we describe the applied methods to construct a
24\micron-based point source catalog derived from the image data of the MIPSGAL
24\micron\ Galactic Plane Survey and the corresponding data products.  The high
quality catalog product contains \catalogtally\ sources, with a total of
\archivetally\ in the full archive catalog.  The source tables include
positional and photometric information derived from the 24\micron\ images,
source quality and confusion flags and counterpart photometry from matched
2MASS, GLIMPSE, and WISE point sources.   Completeness decay data cubes are
constructed at 1\arcmin\ angular resolution that describe the varying
background levels over the MIPSGAL field and the ability to extract sources of
a given magnitude from this background.  The completeness decay cubes are
included in the set of data products.  We present the results of our efforts to
verify the astrometric and photometric calibration of the catalog, and present
several analyses of minor anomalies in these measurements to justify adopted
mitigation strategies.  

\end{abstract}





\section{Introduction}

The MIPSGAL Survey is a Legacy Program of the Spitzer Space Telescope that
imaged the 24 and 70\micron\ emission along the inner disk of the Milky Way
\citep{carey2009}.  These mid-infrared bands are sensitive to the thermal
emission radiated by interstellar dust grains that reside within a broad range
of environments such as the envelopes of evolved stars, circumstellar disks and
infalling envelopes surrounding young stellar objects, HII regions, supernova
remnants, and the extended domains of dense, interstellar clouds.  As a wide
area survey, MIPSGAL is an important component to the infrared-to-millimeter
reconnaissance of the Galaxy, which includes recent, all-sky missions: 2MASS
\citep{skrutskie2006}, WISE \citep{wright2010}, and Planck \citep{planck2011}
as well as surveys targeted along the Galactic plane: GLIMPSE
\citep{churchwell2009}, ATLASGAL \citep{schuller2009}, the Bolocam Galactic
Plane Survey \citep{aguirre2011, ginsburg2013} and the Herschel Infrared
Galactic Plane Survey \citep{molinari2010}.  With its primary 24\micron\
band\footnote{Operational non-linearities in the MIPS 70\micron\ system
resulted in much reduced sensitivities.  The Spitzer 70\micron\ data are
superseded by the Herschel/PACS system and are not included in the MIPSGAL
catalog described in this contribution.}, MIPSGAL provides a critical
wavelength measurement, which links the near infrared data from  2MASS and
GLIMPSE to the far-infrared/submillimeter information for both point sources
and diffuse emission. 

The processed, 24\micron\ MIPSGAL image mosaics have been available since 2008
\citep{carey2009}.  This data product is comprised of flux calibrated FITS
images of 24\micron\ surface brightness with astrometric header information,
images of the surface brightness standard deviations of the coadded data, data
coverage and locations of problematic data.  Each mosaic field (hereafter, a
tile) covers $\sim$ 1$\times$1 deg$^2$ area.  

As much of the measured MIPSGAL 24\micron\ signal resides within an unresolved
component (evolved stars, young stellar objects, compact clusters), a
previously missing yet critical data product is a point source catalog derived
from the image tiles.  The value of a source catalog lies within the uniformity
of the source extraction and photometry algorithms applied to all image data
and the evaluation of source completeness.  The compilation of source
positions, fluxes, flux errors, and completeness limits enables a more
comprehensive, condensed examination of 24\micron\ emitting objects in the
Galaxy.  When merged with photometry from other surveys, one can further select
for certain types of objects based on the shape of the spectral energy
distribution and flux amplitude. 

In this contribution, we describe the construction of a 24\micron-based source
catalog derived from MIPSGAL data.  In \S2, the source extraction and aperture
photometry methods are summarized.  The photometric accuracies, calibration,
and catalog completeness are evaluated in reference to the literature in \S3.
In \S4, we describe the method to derive 24\micron\ source completeness limit
for each MIPSGAL tile.  The columns of the source catalog table are defined in
the Appendix.  

\section{Building the 24\micron\ Point Source Catalog}

Here we describe in detail the methods used in the construction of the
inclusive ``archive'' and high reliability ``catalog'' photometry tables using
the MIPSGAL 24\micron\ image tiles.  In summary, we find compact sources in all
tiles, measure their 24\micron\ photometric properties, merge the tile lists
together, and link the results to external catalogs.  Astrometric systematics
are examined in order to correct calibration offsets by tile and establish
conditions for a confusion flag that is internal to our source list.

\subsection{Source Extraction} 


The MIPSGAL image tile products are extremely uniform integration depth maps of
24\micron\ flux density, but robust point source detection is nontrivial
because of nonuniform background emission across the Milky Way.  For a large
survey such as MIPSGAL, automated data analysis techniques are essential.
However, many automated point source detection techniques produce substantial
numbers of false detections among the filamentary emission structures of the
nebulae surrounding recent star forming events.  Here, we have adopted the IDL
program PhotVis (version 1.10) to robustly identify point-like sources
regardless of the complexity of the background \citep{gutermuth2008}.  

PhotVis employs a modified version of the DAOFIND source detection algorithm
\citep{stetson1987}, as implemented in the IDL Astronomy Users' Library
\citep{landsman1993}.  In summary, the DAOFIND technique involves convolving
each image with a ``sunken'' two dimensional Gaussian function sized to match
the beam size of the observations (for this work, 6.25\arcsec\ full-width at
half-maximum; FWHM).  This convolution concentrates the flux of unresolved
structure into the central pixels of that structure, while large scale
structure effectively convolves to a value near zero.  Ideally, the convolved
image makes stellar sources easy to identify with a simple threshold search.
Unfortunately, numerous false sources are found by this algorithm among
filamentary and other nonuniform structure in the bright nebulosity associated
with the Galactic plane, and regions of star formation more generally
\citep[e.g.][]{megeath2004}.  In PhotVis (v1.10), the standard DAOFIND
algorithm has been enhanced to include empirical estimation of a noise map for
the Gaussian-convolved source detection image.  Specifically, the absolute
value of the convolved image is boxcar median-smoothed with a box size of five
times the FWHM of the point spread function (PSF).  The original
Gaussian-convolved image is then divided by this noise map, effectively
converting the search threshold from a signal-based threshold into an
approximate local signal-to-noise-based threshold.  We use a threshold value of
seven in the local noise map scale, based on considerable testing on MIPS
24\micron\ data of star-forming regions
\citep[e.g.][]{gutermuth2008,gutermuth2009,beerer2010,megeath2012}.  The
resulting algorithm simultaneously achieves excellent sensitivity in dark,
uniform, uncrowded regions of images and automatic adaptation to less sensitive
local conditions, largely mitigating the production of false sources associated
with nebulous structure \citep{gutermuth2008}.  

Once sources are found, their flux is measured using synthetic aperture
photometry via {\it aper.pro} from the IDL Astronomy Users' Library
\citep{landsman1993}.  The MIPSGAL tiles are made of merged observations at a
range of spacecraft rotation angles, thus we chose to use aperture photometry
rather than PSF-fitting photometry due to its computational simplicity and
measurement robustness under that circumstance.  We adopt aperture and inner
and outer sky annulus radii of 6.35\arcsec, 7.62\arcsec, and 17.78\arcsec\,
respectively, and a magnitude zero point of 14.525~mag (Vega standard) for a 1
Digital Number per second (DN/s) source observed at 24\micron\
\citep{gutermuth2008}\footnote{The original \citet{gutermuth2008} magnitude
zero point is 14.6~mag.  We have applied a 0.075~mag reduction to account for
the sightly smaller aperture radius here relative to that work.  Under the
assumption of a `FLUXCONV' value of 0.0447~MJy/sr per DN/s and a Vega flux of
7.17~Jy at the MIPS 24\micron\ channel's isophotal wavelength of 23.68\micron,
our zero point results in an aperture correction factor of 1.63.  The MIPS
Instrument handbook suggests an aperture correction of 1.61 for a 7\arcsec\
aperture, thus our slightly larger correction for our slightly smaller aperture
radius (6.35\arcsec) is consistent.}.  The photometric uncertainty is derived
from calculations of the shot noise in the aperture and shot noise and internal
variance in the sky annulus pixels that are used to compute the background
emission per pixel for subtraction from the aperture flux.  An internal noise
floor of 0.02~mag is enforced to prevent rare data anomalies from yielding
untenable uncertainty estimates. 

Finally, as a characterization of source quality, we compute the FWHM of each
source.  As noted above, calibration of aperture photometry includes a
correction for the finite sampling of the PSF set by the choice of aperture and
sky radii.  If an object is intrinsically resolved beyond the instrument
resolution, then the source would be of relatively poor photometric quality in
our catalog because the aperture correction would be incorrect.  The
measurement algorithm used is entirely empirical, extracting the half of peak
flux radial distance from a cubic spline interpolation of the radial profile
\citep{barth2001}.  We azimuthally average (by median) the radial profile
before running this algorithm to improve measurement success probabilities near
structured nebulosity.

\subsection{Archive Construction\label{arch}}

Once the source lists and photometry have been obtained from all of the
individual tiles, we combine them into a unified survey ``archive'' data
product.  The tiles were constructed with some degree of overlap, thus
duplicate detections near tile edges are common and must be identified and
removed.  Once astrometry systematics were treated (see Section~\ref{astrom}),
a simple angular offset tolerance of 1\arcsec\ is used to identify all
inter-tile duplicates.  For each set, the instance of the source that is
furthest from tile edges is selected to represent that source in the final
combined source list as this maximizes the coverage of the sky annulus and the
surrounding area for the noise map calculation.  The resulting tally of
detections in the final archive that have $<$0.33~mag uncertainty at 24\micron\
is \archivetally.  This requirement is approximately a Signal-to-Noise Ratio
(SNR) of 3, significantly lower than the approximate SNR$>7$ limit mentioned
above for our empirically derived noise map in the source identification
process.  The photometrically determined uncertainty is generally somewhat
higher because it includes photon shot noise.  Ultimately, the photometric SNR
limit is a sensitivity limit, but not where the survey is {\it complete}, as we
will explore in Section~\ref{completeness}, below.  In Figure~\ref{snrandfwhm},
the variations of magnitude uncertainties (top) and FWHM (bottom) with
magnitude for the archive sources are expressed as two-dimensional histograms.
The spread in magnitude uncertainty for a given value of [24] simply reflects
the variation of backgrounds throughout the MIPSGAL field.  


Via automated queries to the Vizier online catalog service, we obtain all of
the 2MASS, GLIMPSE, and WISE sources that fall within each tile.  These are
matched to our MIPSGAL archive such that the closest match within an angular
tolerance of \matchtolerance\ is linked to each 24\micron\ source.  The
matching tallies for each data source are summarized in Table~\ref{matchtbl}.
A counterpart is found in at least one of these catalogs for over 94\% of
sources in the archive.  
To gauge mismatch rates for each catalog, we performed simple Monte Carlo tests
across the entire archive product.  Taking the number of objects within 6.35''
diameter reported for characterization of potential beam contamination for each
archive source (reported in the source table), we compute the mean density of sources
near each object.  Note that this includes the matched source. Thus if this is
a true match, we will be overestimating the field density somewhat (10-50\% for
GLIMPSE and 2MASS, 100\% for WISE, typically).  We then multiply that density
by the area corresponding to the smaller 2'' matching radius to determine a
mean number of contaminators to expect for that source.  Using the mean
contaminator rate, we pull random Poisson deviate numbers of potential
contaminators for each object.  For each object with a non-zero synthetic
contaminator count $n_i$ in a given realization, we draw that number of
uniform, area-weighted deviates (i.e. radius PDF = 2r, per the classic
dartboard problem) and compare the smallest value to the radial separation,
$r_{match}$, of the actual match for the archive source. If the nearest false
source is within $r_{match}$ + 0.1\arcsec\, we count that as a possible
contamination event in the test.  We then integrate contamination counts over
the entire archive, over 1000 trials.
The resulting estimated mismatch rate is $\sim$0.1\% for each catalog (see Table~\ref{matchtbl}). 

Source quality flags are compiled for each source, including the FWHM
(described above), a binary flag to note sky annulus overlap with image edges,
coverage edges, or saturated pixels, and source proximity among nearest
neighbor archive members, in arcseconds.  An internal confusion flag
based on the nearest neighbor distance and the difference in 24\micron\
magnitude between source and neighbor is described in Section~\ref{astrom}.  We
also tabulate the number of objects in each external catalog that fall within
6.35\arcsec\ of the source's centroid position.


\subsection{Astrometric Systematics\label{astrom}}

Initial efforts to incorporate publicly available external catalogs with our
24\micron\ archive revealed systematic offsets in the astrometric calibration
of the MIPSGAL tiles.  These offsets are shown in Figure~\ref{astromfixfig} as
astrometric residuals between the GLIMPSE and 24\micron\ centroid positions
($\Delta RA(i) = [RA_{MG}(i) - RA_{GL}(i)]~cos(Dec_{GL}(i))$; $\Delta Dec(i) =
Dec_{MG}(i) - Dec_{GL}(i)$) for all unconfused matches (specifically, we
require one unique GLIMPSE source within \matchtolerance, and no other GLIMPSE
sources within 6.35\arcsec) in the archive as a function of Galactic Longitude.
Many of these offsets are greater than 0.5\arcsec\ and much larger than the
expected random error between centroid differences.  The bulk of the deviations
can be cured with a constant RA-Dec offset corresponding to the median of the
offsets in each tile.  These tile-specific offsets have been applied to each
tile's source catalog.  Figure~\ref{astromfixfig} shows the residuals after the
application of the offset.  The applied offsets and the final RMS residuals in
RA and Dec for each tile are recorded in Table~\ref{astromfixtab}.  A similar
issue was reported in the Galactic center MIPS coverage in \citet{hinz2009},
and was addressed in a similar manner, using 2MASS for reference astrometry
instead of GLIMPSE.  Additional astrometric systematics internal to many tiles
are present, but treating these would most likely require rebuilding the tiles
from the BCD data products.  

We identified a secondary issue related to astrometry in the archive's nearest
neighbor distance ($d_{NN}$) distribution shown in Figure~\ref{nndist}.  The
functional form of the distribution is approximately log-normal, with a narrow
true normal excess centered on 10\arcsec\ angular separation.  That distance
corresponds to the central radius of the first diffraction ring outside of the
Gaussian core of the MIPS 24\micron\ PSF, suggesting that one of the pair could
be a false identification.  Moreover, such a feature can skew the photometry
and astrometry of faint sources that fall near the feature.  The magnitude
difference between each source and its nearest neighbor in the archive versus
$d_{NN}$ is displayed in Figure~\ref{nndvsmag}a.  The same data are plotted for
those objects without and with GLIMPSE counterparts within 2\arcsec\ in
Figs.~\ref{nndvsmag}b~and~\ref{nndvsmag}c, respectively. 
The distribution of magnitude differences for sources without GLIMPSE
counterparts exhibits clear excess source counts in three distinct locations:
-0.2$<\Delta$[24]$<$0.2 \& $d_{NN}<$8\arcsec, $\Delta$[24]$>$0.8 \&
9\arcsec$<d_{NN}<$11\arcsec, and $\Delta$[24]$>$2.8 \&
25\arcsec$<d_{NN}<$27.5\arcsec.  This excess is further illustrated in
Figure~\ref{nndvsmag}d that shows the ratio of the magnitude differences of
sources without and with GLIMPSE counterparts and normalized by the expected
ratio uncertainty, assuming Poisson counting statistics.  Guided by this
figure, where the grayscale has been set to mark $>$3$\sigma$ regions as black,
we define the conditions for the internal confusion flag.  The
conditions and the source counts affected are listed in
Table~\ref{confuseflag}.  


\subsection{Catalog Construction}

The archive data product is meant to be an inclusive list of 24\micron\
point-like sources extracted from the MIPSGAL survey.  A higher reliability
subset of the archive sources, the ``catalog'' data product, is selected to
mitigate the systematic issues in the archive discussed in
Sections~\ref{arch}~\&~\ref{astrom}.  First, we impose a more stringent
$<$0.20~mag uncertainty requirement (SNR$\sim$5).  Then, we require a confined
range of source FWHM, with thresholds that are flared to a wider range for
dimmer sources to allow for their larger FWHM variance.  These two stricter
limits are drawn in Fig.~\ref{snrandfwhm}, and they yield the vast majority of
the archive objects that get culled from the catalog, roughly evenly divided
between the two constraints.  In addition, we require that the binary confusion
and edge flags must be zero to ensure that these relatively rare instances are
also culled from the catalog.  All of the requirements for catalog inclusion
are listed in Table~\ref{catflg}.  There are \catalogtally\ sources that meet
these more restrictive requirements.  A counterpart in at least one of WISE,
2MASS, or GLIMPSE is found for over 98\% of sources in the catalog, compared to
94\% in the archive.

\section{Archive and Catalog Verification}

The MIPSGAL survey lacks any deep observations of verification fields to enable
direct evaluation of the effectiveness and reliability of our source detection,
astrometry and photometry algorithms, as is often done by large, shallow
surveys such as 2MASS \citep{skrutskie2006}.  Here we perform several analyses
of our methods by comparison to other previous studies and surveys in order to
bootstrap some measures of reliability for extracted sources.

In Figure~\ref{maghist}, we present the magnitude versus uncertainty
distribution and magnitude histogram for the entire survey, as well as for two
regions of the survey that are chosen to demonstrate the extremes in
sensitivity changes set by location within the Galactic plane: the densely
populated regions of the inner bulge and central disk, and the less densely
populated off-plane areas of the wider survey.  We have defined Galactic
coordinate cuts of $|b|<0.5$ and $|l|<10$ for the ``Central Bulge'' region, and
$|b|>0.5$ and $|l|>15$ as the ``Disk, Off-Plane'' zone.  We use these divisions
in several figures through the rest of this paper.  In summary, the one
magnitude relative shift (7 vs 8) in the locations of the peaks of the
magnitude histograms is an initial demonstration of the substantially reduced
sensitivity of the bulge area of the survey relative to the off-plane zone.
With reduced crowding, less bright sources, and less nebulosity, the off-plane
portion of the survey is much more sensitive to fainter objects.

\subsection{Robitaille et al. 2008}

In order to verify the photometric performance and calibration of our source
extraction process, we merged the MIPS 24\micron\ photometry of red sources
provided in \citet{robitaille2008} (R08) with our catalog.  The base image
dataset is the same in both cases, but R08 used the original Spitzer Science
Center pipeline-reduced mosaics for their photometry instead of the enhanced
MIPSGAL-reduced tiles.  Regarding source extraction, they also used PSF fitting
photometry by hand, instead of automated aperture photometry as we have done
here.  Of the \robittally\ red GLIMPSE sources in the R08 catalog,
\robitmipstally\ have reported MIPS 24\micron\ fluxes and uncertainties.
Matches for \robitarchtally\ of those sources are made within the archive
product (97.6\%), and \robitcattally\ matches (92.8\%) with the catalog
product.  The magnitude residuals between the R08 photometry and ours are
plotted in Figure~\ref{robitresid}.  The 2D histogram grayscale shows the
magnitude residuals to R08 matches in the archive, and the contours represent a
similar 2D histogram that uses the catalog product instead.  The mean
zero-point calibration offsets are \robitarchoffset\ and \robitcatoffset, and
the RMS deviations are \robitarchrms\ and \robitcatrms\ for the archive and
catalog products, respectively.  In summary, we find that our photometry agrees
well with the limited photometric sample of R08.

\subsection{Hinz et al. 2009}

As a secondary check, we merged our archive with the MIPS 24\micron\ photometry
of the Galactic center region from \citet{hinz2009} (H09).  As with R08, the
image datasets are the same as ours, but the image data treatment and source
extraction differ.  In this case, the image data were processed with the MIPS
instrument team's DAT software \citep{gordon2005}, and the photometry was
extracted via PSF fitting.  The benefit of this reference catalog over that of
R08 is that it is a complete catalog of 24\micron\ sources from the region in
question, instead of targeted photometry of red 2MASS and GLIMPSE sources
across the entire inner Milky Way.  As such, it is a good test of our
completeness within one of the most challenging parts of the survey.  Of the
\hinzmipstally\ sources in \citet{hinz2009}, we have \hinzarchtally\ and
\hinzcattally\ coincident sources in our archive and catalog products,
respectively.  Obviously, this is a substantial miss rate.  In
Figure~\ref{hinzresid}, we plot the magnitude residuals versus magnitude in the
top plot, demonstrating largely consistent photometry among matches.  Thus
while the photometry appears to agree, the issue of the discrepant sources
demands further characterization.  

In order to fairly examine the sources within a well-covered region, we first
crop both the H09 catalog and our archive to an easily defined common coverage
region of $-3 < l < 4$ and $|b| < 0.5$.  Within this region, we find
\hinzcuttally\ and \hinzcutarch\ sources from the H09 catalog and our archive
product, respectively.  Among those two source lists, \hinzcutmatch\ sources
match.  The bottom panel of Figure~\ref{hinzresid} shows the relative detection
fraction per 1~mag bin among the matched sources (solid line), those found in
our archive but missed by H09 (dot-dashed line), and those missed in our
archive but found in H09 (dashed line) with the common coverage region.  In the
brightest bin, we see a clear deficit of H09 sources.  Generally, those with
marginally detectable peak saturation are rejected by the PSF fitting of H09
but are included in our archive.  The range $1<[24]<6$~mag exhibits consistent
behavior: 70\% matched sources, 10\% H09-only sources, and 20\% H09-missed
archive sources.  At $[24]>6$~mag, the fraction of sources rapidly becomes
dominated by the H09 source counts, as our archive loses completeness
(characterized in detail in Section~\ref{completeness}, below).  We visually
inspected some of the faint H09 source positions in the MIPSGAL tiles and found
that the vast majority of those that we viewed are not apparent in those data.
This effort was sufficient to cement our confidence in our method's omission of
these fainter sources.  Further investigation of the veracity of the faint H09
sources is beyond the scope of this paper.

\subsection{WISE 22\micron}

Finally, the merger of the MIPSGAL archive with the all-sky WISE catalog
enables a check for general agreement between our photometry and the WISE
22\micron\ photometry on a larger sample of objects.  We find that \wisematch\
objects have reported $<$0.33~mag uncertainties in both the MIPSGAL archive and
WISE 22\micron\ catalogs.  We plot the magnitude residuals versus WISE
22\micron\ magnitude in Figure~\ref{wiseresid}.  The median residual for bright
sources ([24]$<$3 mag) in the ``Disk, Off-Plane'' field is \wisemagoffset,
similar to the offset to the MIPSGAL photometry reported in R08 and discussed
above.  The bias toward brighter values in the faint source WISE photometry is
frequently observed in lower relative resolution data, where structured nebular
emission is more likely to contaminate the photometric aperture relative to the
surrounding sky in some sources, resulting in background flux underestimation
and source flux overestimation \citep[e.g.][]{gutermuth2009}.  In this case, the WISE 22\micron\ fluxes of some sources are found to be as much as 3 mag brighter than the MIPSGAL photometry.

%

\section{Completeness Characterization\label{completeness}}


The general means to test the effective sensitivity of a given photometric
survey dataset and a given source extraction and analysis algorithm is to add
false sources to the data and attempt to recover them.  Many papers have
acknowledged spatial variations in such completeness tests, but few have
presented a detailed characterization.  One recent effort to characterize and
treat this effect was performed as part of the analysis of the {\it Spitzer}
survey of the Orion Molecular Clouds \citep{megeath2012}.  That work emphasized
probing locations near where objects of interest, YSOs in this case, have
already been detected.  As with any nearby star-forming region, the 24\micron\
Galactic plane has many areas of bright and structured nebulosity where point
source sensitivity will be reduced.  Any catalog produced from these data would
only be complete with respect to this spatially varying point source
sensitivity, and thus the impact of this effect is important to characterize in
some detail \citep[e.g.][]{gutermuth2005,megeath2012}.  

Many science goals, such as constructing luminosity functions or analyzing
source clustering, demand a spatially unbiased characterization of varying
completeness.  We have mapped this effect and provide it as a companion to the
point source catalog and archive products.  To quantify source completeness, we
have adopted and updated the method described in \citet{gutermuth2005} for this
purpose, at a grid sampling resolution of $1\arcmin\ \times 1\arcmin$.  The
local completeness decay as a function of source flux in each grid cell is
evaluated by performing successive trials of adding and recovering false
sources of varying flux.  Fluxes are sampled in 0.5~mag steps over a typical
range from $0 < [24] < 10$~mag.  Each 1\arcmin\ cell is sampled completely by
adding sources at each position in a 3.125\arcsec\ grid within the cell,
thereby Nyquist sampling the 6.25\arcsec\ FWHM beam width of the MIPS
24\micron\ channel.  The resulting total is $\sim$400 sources per flux step per
cell.  The tally for each flux step and cell is normalized to represent a
fractional completeness.  For each MIPSGAL tile, a data cube of dimensions $60
\times 60 \times 20$ represents the differential completeness
fraction for each cell position within the 1~deg$^2$ tile as a function of the
$\sim$20 flux steps of 0.5~mag.  

In Figure~\ref{goodbadcomp}, we plot examples of the differential completeness
as a function of source flux for two contrasting locations, one with a smooth
and low surface brightness background and the other with a structured and high
surface brightness background.  The low background case demonstrates a clear
increase in sensitivity to faint sources relative to the bright, highly
structured field.  In addition, the rate of decay as a function of source flux
varies between these two examples.  Using the difference between the 20\% and
the 90\% differential completeness limits as an estimator of this effect, there
is a slower completeness decay in the less sensitive area (1.8 vs. 1.4~mag in
the plotted examples).  Despite the potential differences among completeness
decay curve shapes, assigning a completeness value to each source in the
archive is valuable as a convenient indicator of local source sensitivity.  For
each source, the completeness decay curve is extracted from the spatially
nearest position to the source in the completeness cube.  A linear
interpolation of this curve is used to determine the magnitude at which 90\% of
the sources are successfully recovered.  This 90\% differential completeness
limit, named ``diffcomp90'' here, is listed for each archive and catalog
source.  Since some science objectives may require higher or lower completeness
percentages than 90\%, the corresponding limiting magnitude can be derived from
these differential completeness data cubes.    

The correlations of source fluxes in the catalog with their local diffcomp90
values are shown in Figure~\ref{compvflux} as a two-dimensional histogram.  The
most obvious feature of this plot is the strong linear feature where moderate
to bright sources are correlated with their diffcomp90 value such that
[24]$\sim$diffcomp90-1.  This correlation is expected as we have sampled the
completeness at such high spatial resolution relative to the large MIPS
24\micron\ PSF.  Any region of otherwise dark background will have effectively
reduced sensitivity due to the presence of a relatively bright source, and that
sensitivity reduction will be correlated with the flux of that source.  In
addition, Fig.~\ref{compvflux} also shows that a completeness limit is not the
same as a sensitivity limit.  Regions of relatively bright diffcomp90 are
rarely uniform within the sampling area, thus objects considerably fainter than
the completeness limit are often detected.  In contrast, it is unlikely that a
bright source will be found in a region of high diffcomp90 magnitude, as the
very presence of a moderate to bright source reduces the local completeness, as
noted above.


\section{Summary}

We present the results of a full point source extraction from the entire
MIPSGAL 24\micron\ enhanced mosaics of the Milky Way. 

\begin{itemize}
\item Over $1.3 \times 10^6$ sources have been identified, photometered, and characterized for source quality via FWHM and nearest neighbor distance ($d_{NN}$) measurements in our archive product. \\ 
\item The archive source list has been matched with several complementary catalogs from the public archives (2MASS, GLIMPSE, WISE), yielding a substantial new multiple bandpass photometric resource for the community.  Over 94\% of the MIPSGAL sources have a counterpart in at least one of the external catalogs. \\
\item We have used comparisons to these large surveys as well as some MIPSGAL photometry in the literature to evaluate the astrometric and photometric veracity of our archive and examine its completeness. Based on this work, constant astrometric offsets were applied to each tile. \\ 
\item Ideal ranges of source quality measurements were identified from which a high reliability catalog product was constructed.  The catalog is composed of over $9 \times 10^5$ sources that obey the more stringent constraints. \\
\item We measured source detection completeness decay as a function of source flux at 1 square arcminute scale for the entire MIPSGAL 24\micron\ survey.  The data cubes (one for each MIPSGAL tile) resulting from this effort are provided as a companion product to aid in subsequent analysis of the catalog. \\
\item The catalog and non-catalog archive source lists, as well as the completeness decay cubes in FITS format, are hosted and publicly available in the Infrared Science Archive (IRSA) at Caltech's Infrared Processing and Analysis Center (IPAC). 

\end{itemize}



\acknowledgments

The authors thank our referee, Tom Robitaille, for constructive feedback that
improved this manuscript.  The authors gratefully acknowledge support for this
project from NASA ADAP grant NNX13AF08G.  This research has made use of the
NASA/ IPAC Infrared Science Archive, which is operated by the Jet Propulsion
Laboratory, California Institute of Technology, under contract with the
National Aeronautics and Space Administration.  This publication makes use of
data products from the Two Micron All Sky Survey, which is a joint project of
the University of Massachusetts and the Infrared Processing and Analysis
Center/California Institute of Technology, funded by the National Aeronautics
and Space Administration and the National Science Foundation.  This publication
makes use of data products from the Wide-field Infrared Survey Explorer, which
is a joint project of the University of California, Los Angeles, and the Jet
Propulsion Laboratory/California Institute of Technology, funded by the
National Aeronautics and Space Administration.  This research has made use of
the VizieR catalog access tool, CDS, Strasbourg, France.  This work is based on
is based on archival data obtained with the Spitzer Space Telescope, which is
operated by the Jet Propulsion Laboratory, California Institute of Technology
under a contract with NASA.



{\it Facilities:} \facility{Spitzer}.



\appendix

\section{MIPSGAL 24\micron\ Point Source Table Column Reference}

Here we provide a reference listing of the columns delivered in the archive and catalog data product tables as they appear in the tables hosted in the IRSA at IPAC. 

\begin{itemize}
\item l, b, RA, Dec: Galactic Longitude, Latitude, Right Ascension and Declination, in degrees, J2000, ICRS reference. \\
\item Fnu\_XX: Flux Density at the noted bandpass, XX, in mJy. \\
\item sigma\_Fnu\_XX: Flux Density Uncertainty at the noted bandpass, XX, in mJy. \\
\item Mag\_XX: Vega-standard Magnitude at the noted bandpass, XX. \\
\item sigma\_Mag\_XX: Magnitude Uncertainty at the noted bandpass, XX. \\


\item SURVEY\_NAME: Source name from the noted SURVEY (e.g. MIPSGAL, TWOMASS, WISE, or GLIMPSE) point source catalog. \\
\item SURVEY\_COUNT: The number of sources from the noted SURVEY (e.g. TWOMASS, WISE, or GLIMPSE) found within the 6.35\arcsec\ MIPSGAL photometric aperture. \\
\item d\_NN: The angular separation in arcseconds between the source and its nearest neighbor within the MIPSGAL archive product.  \\
\item FWHM: Empirically measured full width at half maximum of the MIPSGAL source, in arcseconds. \\

\item Sky\_24: The background flux density measured in the sky annulus in MJy/sr. \\
\item Comp\_Lim\_Fnu\_24: The 90\% differential completeness limit, in mJy. \\
\item Comp\_Lim\_Mag\_24: The 90\% differential completeness limit, in Vega-standard magnitudes.  Refered to as {\it diffcomp90} in the text.\\
\item Edge\_Flag: A binary flag set to 1 when the aperture overlaps with a masked out area of the MIPSGAL tiles, such as saturated areas or coverage edges. \\
\item Int\_Confuse\_Flag: An integer flag set to 0 if unconfused, or 1, 2, or 3 to denote which of the three confusion criteria in Table~\ref{confuseflag} flagged this source. \\


\end{itemize}




\clearpage

\include{figs}

\clearpage

\include{match_table1}

\include{astrom_table1}

\include{confuse_table1}

\include{cat_table1}

\end{document}

%% file: figs.tex
\begin{figure}
\epsscale{1.0}
\plotone{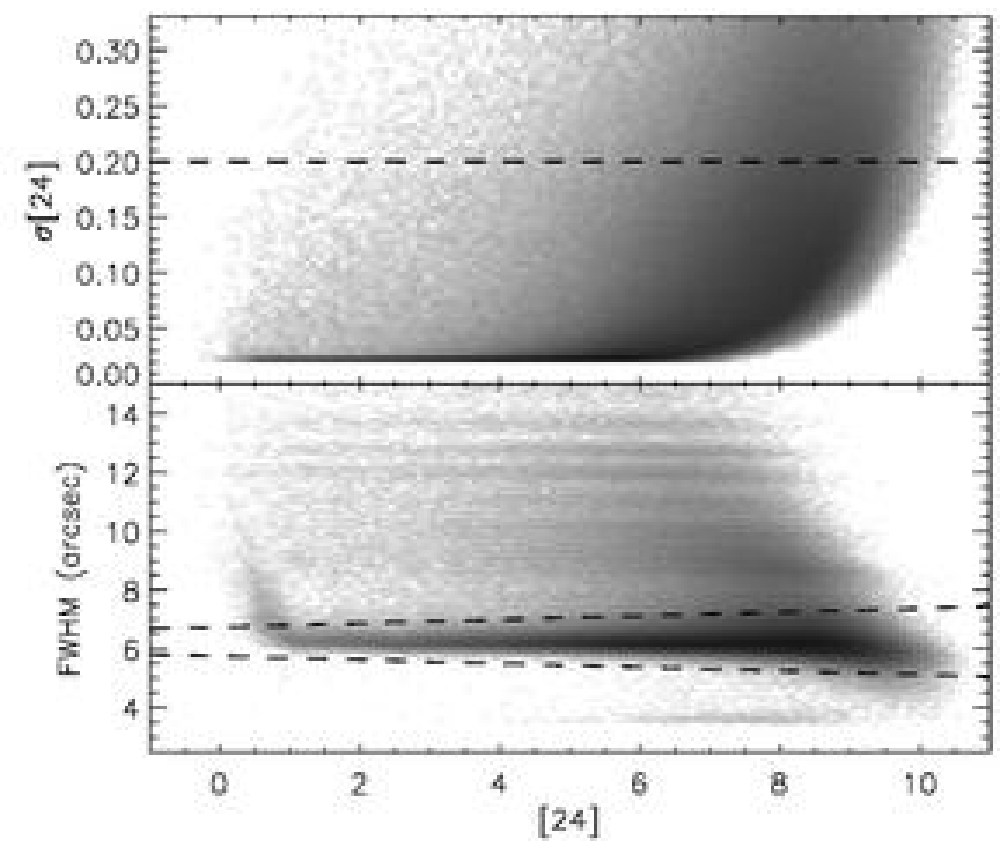}
\caption{Magnitude uncertainty (top) and FWHM (bottom) versus 24\micron\ magnitude for the entire MIPSGAL archive, plotted as a source density map.  The grayscale is inverted log scale, where white is $<$1 and black is $>$$10^4$ sources per bin.  Dashed lines mark the stricter limits imposed on those sources included in the ``catalog'' data product.\label{snrandfwhm}}
\end{figure}

\begin{figure}
\epsscale{1.0}
\plotone{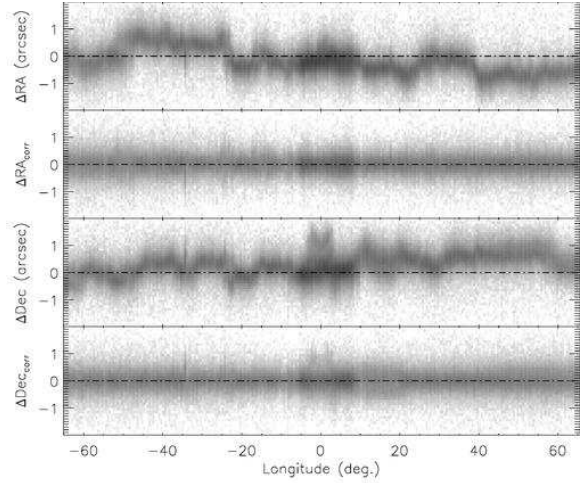}
\caption{Astrometry residuals ($\Delta$RA \& $\Delta$Dec.) to GLIMPSE positions versus Galactic Longitude before and after (``corr'' subscript) correction of systematic astrometric offsets by tile, plotted as a source density map. The grayscale is inverted log scale, where white is $<$1 and black is $>$$10^3$ sources per bin.  Residual spread in the ``corr'' plots is a combination of random variance and systematic variation within tiles that is partially correlated to MIPS scan legs. \label{astromfixfig}}
\end{figure}

\begin{figure}
\epsscale{.80}
\plotone{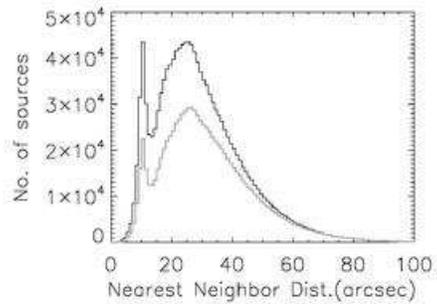}
\caption{Histogram of nearest neighbor distances ($d_{NN}$) in arcseconds for the entire MIPSGAL archive.  The black histogram is computed from all neighbor connections.  For the gray histogram, we have eliminated degenerate entries caused by those object pairs that are each other's nearest neighbors. \label{nndist}}
\end{figure}

\begin{figure}
\epsscale{.80}
\plotone{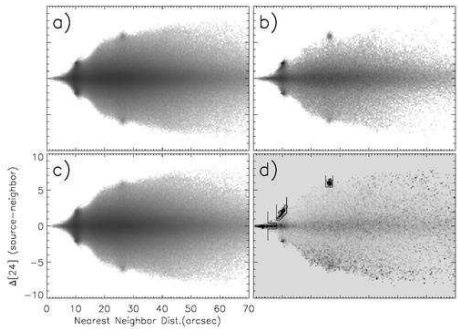}
\caption{Density maps of the magnitude difference between each source and its nearest neighbor versus their separation distance in arcseconds.  In panel a), we plot data for the entire archive.  We split the set between those sources that lack or possess GLIMPSE counterparts in panels b) and c) respectively.  Finally, panel d) shows a map of statistically significant ($>$3$\sigma$ in black) overdensities in panel b) relative to panel c); black lines mark our conditions for flagging a source as ``confused'' and therefore not included in the high quality catalog data product.  The inverted grayscale is logarthmic in panels a), b), \& c), with white for $<$1 source per bin.  Black is $>$$10^3$ sources per bin in panel b) and $>$$10^4$ in a) \& c).  Panel d) is linear scaled from -0.5 to 3, white to black. \label{nndvsmag}}
\end{figure}

\begin{figure}
\epsscale{.80}
\plotone{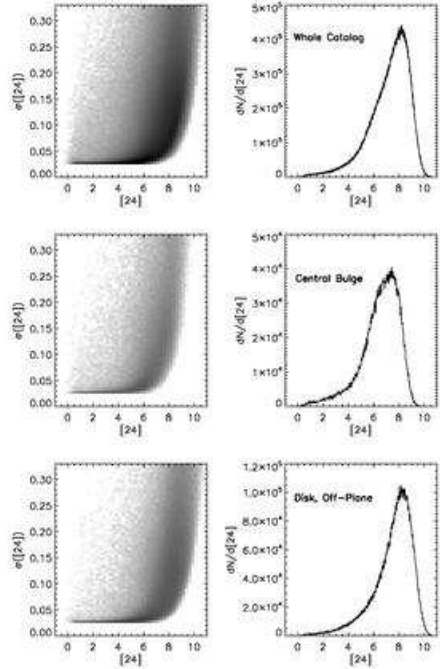}
\caption{Magnitude versus uncertainty source density plots and magnitude histograms for the entire survey (top), and two of samples showing the extremes in sensitivity: the Galactic Center (middle), and the off-plane disk (bottom).  The grayscale is inverted log scale, where white is $<$1 and black is $>$$10^4$ sources per bin.  \label{maghist}}
\end{figure}

\begin{figure}
\epsscale{.80}
\plotone{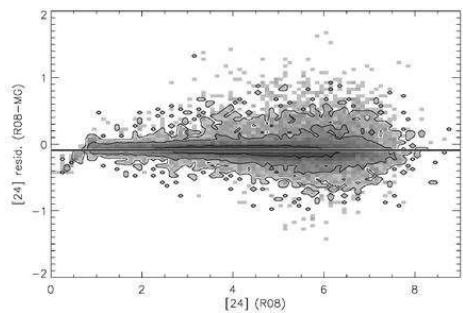}
\caption{Photometry comparison of our MIPSGAL archive to the MIPSGAL photometry reported in \citet{robitaille2008}.  Our calibration offset estimate is plotted as a solid line.  The density map is inverted log scale, with white and black levels set to 1 and $10^3$ sources per bin, respectively.  Contours are plotted at 0.5, 5, and 50 sources per bin. \label{robitresid}}
\end{figure}

\begin{figure}
\epsscale{.80}
\plotone{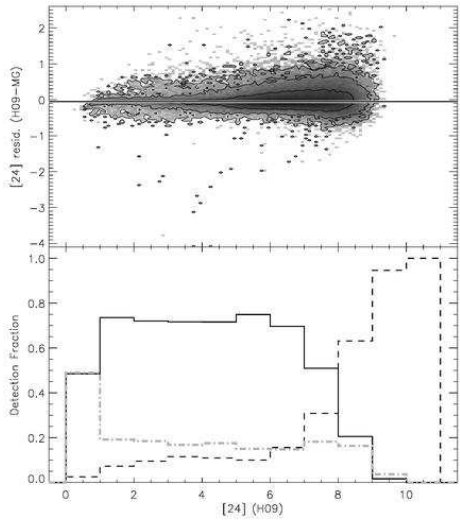}
\caption{Photometry and apparent completeness comparison of our MIPSGAL archive to the MIPSGAL photometry reported in \citet{hinz2009}.  The top panel is similar to Fig.~\ref{robitresid}, above, and the source density scaling and contour levels are identical to that plot. Our calibration offset estimate is plotted as a solid line.  The bottom plot contains the fraction of sources detected in our MIPSGAL archive only (gray, dot-dashed), the H09 catalog only (black, dashed), and both data sources (black, solid), as a function of magnitude. \label{hinzresid}}
\end{figure}

\begin{figure}
\epsscale{.70}
\plotone{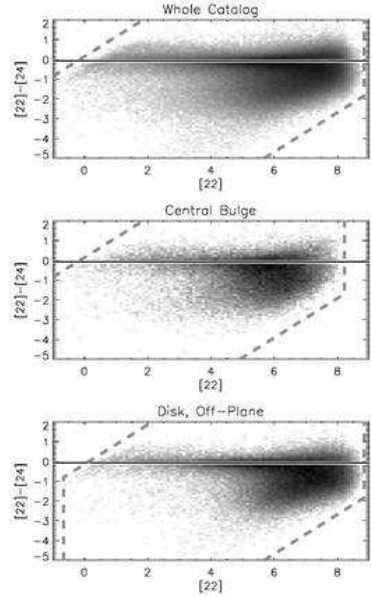}
\caption{Photometry comparison to those MIPSGAL archive sources with WISE 22\micron\ counterparts with $\sigma < 0.33$~mag, plotted as a source density map.  The grayscale is inverted log scale, with white and black levels set to $<$1 and $>$$10^3$ sources per bin, respectively.  Our estimate of the offset calibration is plotted as a solid line.  The approximate local saturation and sensitivity limits on the data space are marked with gray dashed lines.\label{wiseresid}}
\end{figure}



\begin{figure}
\epsscale{.80}
\plotone{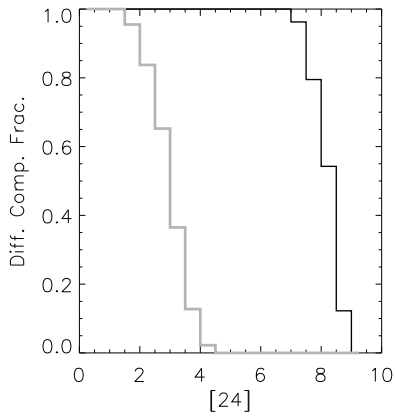}
\caption{A demonstration of two differential completeness fraction decay curves found in extremes of environment.  The gray plot is taken from a location of bright, structured background emission, and its 90\% differential completeness limit is [24]=1.98 mag.  The black plot is taken from a dim background locale, and thus it has a considerably more sensitive 90\% differential completeness limit of [24]=7.44 mag. \label{goodbadcomp}}
\end{figure}

\begin{figure}
\epsscale{.80}
\plotone{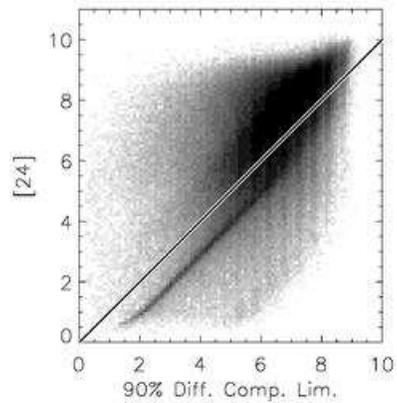}
\caption{A comparison plot of catalog source fluxes to their nearest associated 90\% differential completeness limit value, plotted as a source density map.  The grayscale is identical to Fig.~\ref{wiseresid}.  Overlaid is a simple one-to-one line, for reference.  The most notable feature is the diagonal linear structure at moderate to bright magnitudes, caused by the wings of the PSF these brighter sources directly creating a corresponding decay in the local sensitivity.  Another feature to note is that the 90\% differential completeness limit is not a single source sensitivity limit; sources can be detected to considerably dimmer values than this limit. \label{compvflux}}
\end{figure}

%% file: match_table1.tex
\begin{deluxetable}{cccc}
\tabletypesize{\scriptsize}
\tablecaption{MIPSGAL Catalog Match Summary\label{matchtbl}}
\tablewidth{0pt}
\tablehead{\colhead{Data Source Name} & \colhead{Archive Matches} & \colhead{Catalog Matches} & \colhead{Estimated Archive Mismatches}}
\startdata
Internal & 1,353,228 (100\%) &  933,818 (100\%) & ... \\
2MASS & 1,199,931 (88.67\%) &  880,168 (94.25\%) & $953 \pm 32$ (0.08\%) \\
GLIMPSE & 1,217,143 (89.94\%) &  867,800 (92.93\%) & $1240 \pm 37$ (0.10\%) \\
WISE & 1,138,070 (84.10\%) &  855,725 (91.64\%) & $777 \pm 29$ (0.07\%) \\
Any Match & 1,281,946 (94.73\%) &  918,966 (98.41\%) & ... \\
\enddata
\end{deluxetable}

%% file: astrom_table1.tex
\begin{deluxetable}{ccccc}
\tabletypesize{\scriptsize}
\tablecaption{MIPSGAL Tile Astrometry Offsets and Residuals Summary\label{astromfixtab}}
\tablewidth{0pt}
\tablehead{\colhead{Tile Name} & \colhead{$\Delta$R.A.} & \colhead{$\Delta$Dec.} & \colhead{$\sigma$R.A.} & \colhead{$\sigma$Dec.}}
\startdata
MG0000n005 &  0.123 &  0.296 &  0.412 &  0.528 \\ 
MG0000n015 & -0.201 &  0.065 &  0.293 &  0.316 \\ 
MG0000p005 &  0.011 &  0.583 &  0.379 &  0.499 \\ 
MG0000p015 & -0.263 &  0.274 &  0.315 &  0.334 \\ 
MG0010n005 &  0.105 &  0.410 &  0.410 &  0.575 \\ 
MG0010n015 & -0.217 &  0.118 &  0.301 &  0.321 \\ 
MG0010n025 & -0.176 &  0.056 &  0.302 &  0.327 \\ 
MG0010p005 &  0.122 &  0.625 &  0.362 &  0.513 \\ 
MG0010p015 & -0.167 &  0.233 &  0.299 &  0.324 \\ 
MG0010p025 & -0.229 &  0.089 &  0.316 &  0.343 \\ 
MG0020n005 &  0.005 &  0.160 &  0.376 &  0.598 \\ 
MG0020n015 & -0.187 &  0.066 &  0.313 &  0.311 \\ 
MG0020n025 & -0.127 & -0.039 &  0.313 &  0.326 \\ 
MG0020p005 & -0.009 &  0.679 &  0.338 &  0.521 \\ 
MG0020p015 & -0.054 &  0.183 &  0.301 &  0.316 \\ 
MG0020p025 & -0.207 &  0.242 &  0.311 &  0.325 \\ 
MG0030n005 & -0.121 &  0.018 &  0.377 &  0.459 \\ 
MG0030n015 & -0.204 &  0.063 &  0.306 &  0.311 \\ 
MG0030n025 & -0.142 & -0.040 &  0.345 &  0.338 \\ 
MG0030p005 &  0.034 &  0.402 &  0.335 &  0.601 \\ 
MG0030p015 & -0.046 &  0.096 &  0.317 &  0.325 \\ 
MG0030p025 & -0.143 &  0.149 &  0.316 &  0.340 \\ 
MG0040n005 & -0.287 &  0.060 &  0.367 &  0.382 \\ 
MG0040n015 & -0.212 &  0.021 &  0.307 &  0.318 \\ 
MG0040n025 & -0.140 & -0.006 &  0.318 &  0.322 \\ 
MG0040p005 & -0.045 &  0.009 &  0.344 &  0.461 \\ 
MG0040p015 & -0.092 &  0.078 &  0.319 &  0.327 \\ 
MG0040p025 & -0.211 &  0.151 &  0.351 &  0.336 \\ 
MG0050n005 & -0.402 &  0.095 &  0.345 &  0.344 \\ 
MG0050n015 & -0.227 &  0.007 &  0.303 &  0.314 \\ 
MG0050n025 & -0.125 &  0.028 &  0.333 &  0.327 \\ 
MG0050p005 & -0.237 &  0.102 &  0.361 &  0.378 \\ 
MG0050p015 & -0.071 &  0.065 &  0.309 &  0.317 \\ 
MG0050p025 & -0.147 &  0.153 &  0.331 &  0.329 \\ 
MG0060n005 & -0.390 &  0.161 &  0.390 &  0.383 \\ 
MG0060n015 & -0.204 &  0.034 &  0.358 &  0.351 \\ 
MG0060n025 & -0.108 &  0.096 &  0.333 &  0.318 \\ 
MG0060p005 & -0.411 &  0.094 &  0.351 &  0.333 \\ 
MG0060p015 & -0.086 &  0.068 &  0.307 &  0.310 \\ 
MG0060p025 & -0.081 &  0.120 &  0.303 &  0.327 \\ 
MG0070n005 & -0.291 &  0.264 &  0.372 &  0.376 \\ 
MG0070n015 & -0.157 &  0.116 &  0.322 &  0.331 \\ 
MG0070n025 & -0.082 &  0.120 &  0.338 &  0.342 \\ 
MG0070p005 & -0.448 &  0.119 &  0.346 &  0.350 \\ 
MG0070p015 & -0.115 &  0.007 &  0.351 &  0.342 \\ 
MG0070p025 & -0.092 &  0.099 &  0.312 &  0.331 \\ 
MG0080n005 & -0.203 &  0.305 &  0.338 &  0.397 \\ 
MG0080n015 & -0.118 &  0.181 &  0.312 &  0.334 \\ 
MG0080n025 & -0.067 &  0.187 &  0.324 &  0.335 \\ 
MG0080p005 & -0.400 &  0.205 &  0.358 &  0.363 \\ 
MG0080p015 & -0.168 & -0.032 &  0.385 &  0.397 \\ 
MG0080p025 & -0.070 &  0.020 &  0.296 &  0.326 \\ 
MG0090n005 & -0.211 &  0.410 &  0.333 &  0.423 \\ 
MG0090n015 & -0.135 &  0.317 &  0.330 &  0.335 \\ 
MG0090p005 & -0.269 &  0.294 &  0.331 &  0.387 \\ 
MG0090p025 & -0.091 & -0.094 &  0.307 &  0.340 \\ 
MG0100n005 & -0.329 &  0.593 &  0.355 &  0.456 \\ 
MG0100p005 & -0.250 &  0.412 &  0.328 &  0.424 \\ 
MG0110n005 & -0.470 &  0.707 &  0.357 &  0.453 \\ 
MG0110p005 & -0.388 &  0.564 &  0.346 &  0.448 \\ 
MG0120n005 & -0.537 &  0.556 &  0.375 &  0.458 \\ 
MG0120p005 & -0.426 &  0.635 &  0.329 &  0.434 \\ 
MG0130n005 & -0.532 &  0.450 &  0.381 &  0.444 \\ 
MG0130p005 & -0.448 &  0.571 &  0.343 &  0.433 \\ 
MG0140n005 & -0.493 &  0.366 &  0.377 &  0.445 \\ 
MG0140p005 & -0.457 &  0.458 &  0.359 &  0.433 \\ 
MG0150n005 & -0.420 &  0.358 &  0.377 &  0.451 \\ 
MG0150p005 & -0.460 &  0.383 &  0.323 &  0.414 \\ 
MG0160n005 & -0.427 &  0.377 &  0.353 &  0.429 \\ 
MG0160p005 & -0.444 &  0.385 &  0.325 &  0.403 \\ 
MG0170n005 & -0.449 &  0.330 &  0.374 &  0.437 \\ 
MG0170p005 & -0.356 &  0.337 &  0.381 &  0.441 \\ 
MG0180n005 & -0.409 &  0.396 &  0.373 &  0.424 \\ 
MG0180p005 & -0.389 &  0.338 &  0.348 &  0.423 \\ 
MG0190n005 & -0.602 &  0.485 &  0.385 &  0.428 \\ 
MG0190p005 & -0.454 &  0.445 &  0.351 &  0.432 \\ 
MG0200n005 & -0.632 &  0.553 &  0.372 &  0.401 \\ 
MG0200p005 & -0.556 &  0.569 &  0.355 &  0.419 \\ 
MG0210n005 & -0.698 &  0.572 &  0.375 &  0.364 \\ 
MG0210p005 & -0.633 &  0.573 &  0.349 &  0.410 \\ 
MG0220n005 & -0.561 &  0.459 &  0.394 &  0.381 \\ 
MG0220p005 & -0.768 &  0.574 &  0.372 &  0.373 \\ 
MG0230n005 & -0.507 &  0.439 &  0.388 &  0.383 \\ 
MG0230p005 & -0.646 &  0.506 &  0.391 &  0.377 \\ 
MG0240n005 & -0.514 &  0.508 &  0.394 &  0.379 \\ 
MG0240p005 & -0.405 &  0.400 &  0.383 &  0.397 \\ 
MG0250n005 & -0.306 &  0.474 &  0.387 &  0.396 \\ 
MG0250p005 & -0.217 &  0.291 &  0.381 &  0.385 \\ 
MG0260n005 & -0.188 &  0.309 &  0.390 &  0.382 \\ 
MG0260p005 & -0.214 &  0.345 &  0.371 &  0.381 \\ 
MG0270n005 & -0.117 &  0.271 &  0.397 &  0.381 \\ 
MG0270p005 & -0.213 &  0.340 &  0.411 &  0.393 \\ 
MG0280n005 & -0.079 &  0.170 &  0.380 &  0.369 \\ 
MG0280p005 & -0.112 &  0.284 &  0.413 &  0.401 \\ 
MG0290n005 & -0.054 &  0.223 &  0.368 &  0.405 \\ 
MG0290p005 & -0.035 &  0.169 &  0.412 &  0.384 \\ 
MG0300n005 & -0.117 &  0.384 &  0.374 &  0.403 \\ 
MG0300p005 & -0.090 &  0.162 &  0.363 &  0.388 \\ 
MG0310n005 & -0.130 &  0.512 &  0.405 &  0.422 \\ 
MG0310p005 & -0.152 &  0.328 &  0.407 &  0.415 \\ 
MG0320n005 & -0.084 &  0.597 &  0.358 &  0.388 \\ 
MG0320p005 & -0.179 &  0.418 &  0.348 &  0.395 \\ 
MG0330n005 & -0.063 &  0.637 &  0.358 &  0.402 \\ 
MG0330p005 & -0.117 &  0.457 &  0.353 &  0.393 \\ 
MG0340n005 & -0.131 &  0.692 &  0.378 &  0.397 \\ 
MG0340p005 & -0.117 &  0.528 &  0.361 &  0.381 \\ 
MG0350n005 & -0.231 &  0.664 &  0.364 &  0.400 \\ 
MG0350p005 & -0.094 &  0.573 &  0.350 &  0.389 \\ 
MG0360n005 & -0.211 &  0.607 &  0.355 &  0.374 \\ 
MG0360p005 & -0.150 &  0.586 &  0.358 &  0.374 \\ 
MG0370n005 & -0.245 &  0.556 &  0.376 &  0.396 \\ 
MG0370p005 & -0.225 &  0.587 &  0.354 &  0.380 \\ 
MG0380n005 & -0.323 &  0.523 &  0.381 &  0.394 \\ 
MG0380p005 & -0.302 &  0.587 &  0.363 &  0.371 \\ 
MG0390n005 & -0.503 &  0.552 &  0.367 &  0.418 \\ 
MG0390p005 & -0.586 &  0.718 &  0.396 &  0.394 \\ 
MG0400n005 & -0.829 &  0.637 &  0.418 &  0.389 \\ 
MG0400p005 & -0.728 &  0.744 &  0.342 &  0.390 \\ 
MG0410n005 & -0.858 &  0.615 &  0.377 &  0.387 \\ 
MG0410p005 & -0.762 &  0.686 &  0.378 &  0.385 \\ 
MG0420n005 & -0.730 &  0.600 &  0.378 &  0.401 \\ 
MG0420p005 & -0.844 &  0.673 &  0.379 &  0.387 \\ 
MG0430n005 & -0.707 &  0.647 &  0.381 &  0.395 \\ 
MG0430p005 & -0.848 &  0.674 &  0.369 &  0.370 \\ 
MG0440n005 & -0.657 &  0.640 &  0.365 &  0.399 \\ 
MG0440p005 & -0.746 &  0.576 &  0.384 &  0.386 \\ 
MG0450n005 & -0.642 &  0.694 &  0.361 &  0.397 \\ 
MG0450p005 & -0.718 &  0.583 &  0.351 &  0.397 \\ 
MG0460n005 & -0.570 &  0.663 &  0.366 &  0.415 \\ 
MG0460p005 & -0.711 &  0.588 &  0.373 &  0.422 \\ 
MG0470n005 & -0.621 &  0.645 &  0.329 &  0.394 \\ 
MG0470p005 & -0.579 &  0.682 &  0.376 &  0.404 \\ 
MG0480n005 & -0.673 &  0.678 &  0.328 &  0.370 \\ 
MG0480p005 & -0.598 &  0.678 &  0.367 &  0.398 \\ 
MG0490n005 & -0.731 &  0.653 &  0.432 &  0.460 \\ 
MG0490p005 & -0.660 &  0.622 &  0.378 &  0.420 \\ 
MG0500n005 & -0.667 &  0.650 &  0.415 &  0.412 \\ 
MG0500p005 & -0.768 &  0.670 &  0.372 &  0.403 \\ 
MG0510n005 & -0.740 &  0.640 &  0.372 &  0.399 \\ 
MG0510p005 & -0.728 &  0.707 &  0.374 &  0.427 \\ 
MG0520n005 & -0.744 &  0.694 &  0.354 &  0.408 \\ 
MG0520p005 & -0.737 &  0.672 &  0.393 &  0.424 \\ 
MG0530n005 & -0.637 &  0.713 &  0.351 &  0.406 \\ 
MG0530p005 & -0.813 &  0.649 &  0.383 &  0.413 \\ 
MG0540n005 & -0.728 &  0.803 &  0.386 &  0.399 \\ 
MG0540p005 & -0.857 &  0.709 &  0.415 &  0.424 \\ 
MG0550n005 & -0.635 &  0.751 &  0.352 &  0.408 \\ 
MG0550p005 & -0.775 &  0.698 &  0.372 &  0.422 \\ 
MG0560n005 & -0.579 &  0.639 &  0.364 &  0.406 \\ 
MG0560p005 & -0.644 &  0.676 &  0.363 &  0.404 \\ 
MG0570n005 & -0.592 &  0.496 &  0.353 &  0.425 \\ 
MG0570p005 & -0.607 &  0.746 &  0.333 &  0.389 \\ 
MG0580n005 & -0.660 &  0.367 &  0.328 &  0.400 \\ 
MG0580p005 & -0.618 &  0.733 &  0.378 &  0.389 \\ 
MG0590n005 & -0.675 &  0.278 &  0.371 &  0.408 \\ 
MG0590p005 & -0.600 &  0.565 &  0.349 &  0.434 \\ 
MG0600n005 & -0.604 &  0.237 &  0.380 &  0.414 \\ 
MG0600p005 & -0.635 &  0.360 &  0.390 &  0.419 \\ 
MG0610n005 & -0.579 &  0.183 &  0.374 &  0.380 \\ 
MG0610p005 & -0.639 &  0.342 &  0.348 &  0.390 \\ 
MG0620n005 & -0.611 &  0.203 &  0.320 &  0.381 \\ 
MG0620p005 & -0.621 &  0.287 &  0.352 &  0.395 \\ 
MG0630n005 & -0.540 &  0.188 &  0.369 &  0.397 \\ 
MG0630p005 & -0.580 &  0.269 &  0.369 &  0.406 \\ 
MG0640n005 & -0.485 &  0.236 &  0.340 &  0.395 \\ 
MG0640p005 & -0.570 &  0.172 &  0.347 &  0.394 \\ 
MG0650n005 & -0.457 &  0.262 &  0.351 &  0.367 \\ 
MG0650p005 & -0.519 &  0.185 &  0.363 &  0.403 \\ 
MG0660n005 & -0.411 &  0.244 &  0.360 &  0.467 \\ 
MG0660p005 & -0.511 &  0.195 &  0.376 &  0.359 \\ 
MG2950n005 & -0.156 & -0.362 &  0.445 &  0.430 \\ 
MG2950p005 & -0.199 & -0.464 &  0.405 &  0.402 \\ 
MG2960n005 & -0.208 & -0.246 &  0.421 &  0.432 \\ 
MG2960p005 & -0.085 & -0.432 &  0.380 &  0.334 \\ 
MG2970n005 & -0.250 & -0.019 &  0.414 &  0.403 \\ 
MG2970p005 &  0.004 & -0.424 &  0.388 &  0.329 \\ 
MG2980n005 & -0.263 & -0.007 &  0.437 &  0.409 \\ 
MG2980p005 & -0.011 & -0.420 &  0.379 &  0.377 \\ 
MG2990n005 & -0.312 & -0.059 &  0.431 &  0.405 \\ 
MG2990p005 & -0.172 & -0.251 &  0.444 &  0.445 \\ 
MG3000n005 & -0.356 &  0.041 &  0.407 &  0.379 \\ 
MG3000p005 & -0.308 & -0.093 &  0.409 &  0.395 \\ 
MG3010n005 & -0.340 &  0.099 &  0.407 &  0.339 \\ 
MG3010p005 & -0.320 &  0.036 &  0.418 &  0.358 \\ 
MG3020n005 & -0.276 &  0.103 &  0.410 &  0.369 \\ 
MG3020p005 & -0.311 &  0.110 &  0.399 &  0.372 \\ 
MG3030n005 & -0.226 &  0.023 &  0.430 &  0.394 \\ 
MG3030p005 & -0.345 &  0.038 &  0.394 &  0.369 \\ 
MG3040n005 & -0.122 & -0.057 &  0.411 &  0.385 \\ 
MG3040p005 & -0.250 &  0.010 &  0.402 &  0.360 \\ 
MG3050n005 & -0.068 & -0.056 &  0.432 &  0.401 \\ 
MG3050p005 & -0.173 & -0.027 &  0.426 &  0.397 \\ 
MG3060n005 & -0.127 & -0.038 &  0.404 &  0.354 \\ 
MG3060p005 &  0.013 & -0.155 &  0.420 &  0.385 \\ 
MG3070n005 & -0.128 & -0.063 &  0.379 &  0.380 \\ 
MG3070p005 &  0.056 & -0.263 &  0.386 &  0.351 \\ 
MG3080n005 &  0.022 & -0.173 &  0.443 &  0.396 \\ 
MG3080p005 & -0.000 & -0.295 &  0.393 &  0.368 \\ 
MG3090n005 &  0.272 & -0.168 &  0.481 &  0.409 \\ 
MG3090p005 & -0.114 & -0.290 &  0.414 &  0.372 \\ 
MG3100n005 &  0.493 & -0.035 &  0.455 &  0.386 \\ 
MG3100p005 & -0.081 & -0.288 &  0.413 &  0.363 \\ 
MG3110n005 &  0.638 &  0.059 &  0.430 &  0.409 \\ 
MG3110p005 &  0.099 & -0.209 &  0.482 &  0.432 \\ 
MG3120n005 &  0.659 &  0.038 &  0.439 &  0.400 \\ 
MG3120p005 &  0.348 & -0.161 &  0.499 &  0.402 \\ 
MG3130n005 &  0.664 &  0.021 &  0.403 &  0.363 \\ 
MG3130p005 &  0.604 & -0.012 &  0.432 &  0.393 \\ 
MG3140n005 &  0.687 &  0.077 &  0.394 &  0.360 \\ 
MG3140p005 &  0.750 &  0.102 &  0.399 &  0.378 \\ 
MG3150n005 &  0.665 &  0.202 &  0.379 &  0.377 \\ 
MG3150p005 &  0.697 &  0.231 &  0.390 &  0.360 \\ 
MG3160n005 &  0.633 &  0.368 &  0.373 &  0.390 \\ 
MG3160p005 &  0.597 &  0.255 &  0.370 &  0.350 \\ 
MG3170n005 &  0.577 &  0.405 &  0.414 &  0.411 \\ 
MG3170p005 &  0.578 &  0.207 &  0.434 &  0.380 \\ 
MG3180n005 &  0.586 &  0.302 &  0.436 &  0.430 \\ 
MG3180p005 &  0.525 &  0.314 &  0.362 &  0.360 \\ 
MG3190n005 &  0.646 &  0.276 &  0.404 &  0.397 \\ 
MG3190p005 &  0.446 &  0.332 &  0.364 &  0.364 \\ 
MG3200n005 &  0.658 &  0.268 &  0.449 &  0.386 \\ 
MG3200p005 &  0.591 &  0.351 &  0.410 &  0.384 \\ 
MG3210n005 &  0.515 &  0.354 &  0.404 &  0.390 \\ 
MG3210p005 &  0.755 &  0.380 &  0.396 &  0.356 \\ 
MG3220n005 &  0.441 &  0.380 &  0.369 &  0.368 \\ 
MG3220p005 &  0.599 &  0.435 &  0.393 &  0.362 \\ 
MG3230n005 &  0.413 &  0.194 &  0.378 &  0.398 \\ 
MG3230p005 &  0.418 &  0.350 &  0.370 &  0.361 \\ 
MG3240n005 &  0.443 &  0.087 &  0.365 &  0.354 \\ 
MG3240p005 &  0.389 &  0.229 &  0.379 &  0.366 \\ 
MG3250n005 &  0.473 &  0.049 &  0.371 &  0.374 \\ 
MG3250p005 &  0.436 &  0.168 &  0.400 &  0.409 \\ 
MG3260n005 &  0.248 &  0.250 &  0.430 &  0.441 \\ 
MG3260p005 &  0.363 &  0.081 &  0.414 &  0.452 \\ 
MG3270n005 &  0.503 &  0.168 &  0.433 &  0.432 \\ 
MG3270p005 &  0.424 &  0.003 &  0.397 &  0.395 \\ 
MG3280n005 &  0.481 &  0.283 &  0.404 &  0.410 \\ 
MG3280p005 &  0.453 &  0.112 &  0.400 &  0.409 \\ 
MG3290n005 &  0.421 &  0.385 &  0.351 &  0.377 \\ 
MG3290p005 &  0.435 &  0.218 &  0.383 &  0.400 \\ 
MG3300n005 &  0.347 &  0.408 &  0.348 &  0.377 \\ 
MG3300p005 &  0.448 &  0.326 &  0.370 &  0.391 \\ 
MG3310n005 &  0.343 &  0.373 &  0.407 &  0.424 \\ 
MG3310p005 &  0.426 &  0.382 &  0.390 &  0.379 \\ 
MG3320n005 &  0.381 &  0.339 &  0.416 &  0.402 \\ 
MG3320p005 &  0.352 &  0.420 &  0.383 &  0.394 \\ 
MG3330n005 &  0.462 &  0.311 &  0.481 &  0.467 \\ 
MG3330p005 &  0.353 &  0.391 &  0.399 &  0.384 \\ 
MG3340n005 &  0.577 &  0.306 &  0.396 &  0.389 \\ 
MG3340p005 &  0.404 &  0.315 &  0.396 &  0.377 \\ 
MG3350n005 &  0.391 &  0.130 &  0.515 &  0.415 \\ 
MG3350p005 &  0.520 &  0.360 &  0.369 &  0.356 \\ 
MG3360n005 & -0.114 & -0.116 &  0.446 &  0.409 \\ 
MG3360p005 &  0.419 &  0.418 &  0.434 &  0.420 \\ 
MG3370n005 & -0.195 & -0.370 &  0.379 &  0.403 \\ 
MG3370p005 &  0.171 &  0.139 &  0.477 &  0.450 \\ 
MG3380n005 & -0.330 & -0.230 &  0.389 &  0.407 \\ 
MG3380p005 & -0.309 &  0.078 &  0.399 &  0.400 \\ 
MG3390n005 & -0.354 & -0.128 &  0.374 &  0.366 \\ 
MG3390p005 & -0.410 & -0.060 &  0.361 &  0.371 \\ 
MG3400n005 & -0.364 & -0.142 &  0.398 &  0.399 \\ 
MG3400p005 & -0.384 & -0.078 &  0.381 &  0.371 \\ 
MG3410n005 & -0.385 & -0.138 &  0.372 &  0.381 \\ 
MG3410p005 & -0.392 & -0.160 &  0.338 &  0.333 \\ 
MG3420n005 & -0.344 & -0.128 &  0.346 &  0.355 \\ 
MG3420p005 & -0.391 & -0.188 &  0.351 &  0.363 \\ 
MG3430n005 & -0.112 &  0.008 &  0.394 &  0.369 \\ 
MG3430p005 & -0.330 & -0.088 &  0.371 &  0.365 \\ 
MG3440n005 &  0.000 &  0.186 &  0.396 &  0.353 \\ 
MG3440p005 & -0.332 & -0.023 &  0.348 &  0.351 \\ 
MG3450n005 & -0.013 &  0.297 &  0.380 &  0.376 \\ 
MG3450p005 & -0.098 &  0.167 &  0.383 &  0.371 \\ 
MG3460n005 & -0.099 &  0.209 &  0.403 &  0.377 \\ 
MG3460p005 &  0.011 &  0.170 &  0.383 &  0.383 \\ 
MG3470n005 & -0.250 &  0.158 &  0.366 &  0.341 \\ 
MG3470p005 & -0.097 &  0.211 &  0.381 &  0.360 \\ 
MG3480n005 & -0.249 &  0.217 &  0.329 &  0.344 \\ 
MG3480p005 & -0.181 &  0.219 &  0.366 &  0.358 \\ 
MG3490n005 & -0.260 &  0.201 &  0.351 &  0.365 \\ 
MG3490p005 & -0.201 &  0.197 &  0.355 &  0.365 \\ 
MG3500n005 & -0.313 &  0.174 &  0.349 &  0.364 \\ 
MG3500p005 & -0.253 &  0.159 &  0.354 &  0.363 \\ 
MG3510n005 & -0.377 &  0.173 &  0.374 &  0.359 \\ 
MG3510p005 & -0.293 &  0.135 &  0.418 &  0.368 \\ 
MG3520n005 & -0.332 &  0.104 &  0.375 &  0.404 \\ 
MG3520n025 & -0.002 &  1.089 &  0.311 &  0.338 \\ 
MG3520p005 & -0.351 &  0.172 &  0.382 &  0.351 \\ 
MG3520p015 & -0.230 &  0.007 &  0.317 &  0.345 \\ 
MG3530n005 & -0.239 & -0.049 &  0.357 &  0.361 \\ 
MG3530n015 & -0.125 & -0.061 &  0.309 &  0.359 \\ 
MG3530p005 & -0.346 &  0.131 &  0.396 &  0.397 \\ 
MG3530p015 & -0.235 &  0.077 &  0.407 &  0.368 \\ 
MG3540n005 & -0.276 & -0.070 &  0.349 &  0.343 \\ 
MG3540n015 & -0.119 & -0.071 &  0.333 &  0.371 \\ 
MG3540p005 & -0.328 & -0.025 &  0.337 &  0.359 \\ 
MG3540p015 & -0.299 &  0.068 &  0.322 &  0.326 \\ 
MG3550n005 & -0.330 & -0.051 &  0.346 &  0.346 \\ 
MG3550n015 & -0.296 & -0.020 &  0.307 &  0.308 \\ 
MG3550n025 & -0.263 & -0.003 &  0.326 &  0.341 \\ 
MG3550p005 & -0.285 & -0.043 &  0.349 &  0.349 \\ 
MG3550p015 & -0.219 & -0.083 &  0.291 &  0.306 \\ 
MG3550p025 & -0.203 & -0.097 &  0.299 &  0.318 \\ 
MG3560n005 & -0.180 &  0.120 &  0.370 &  0.435 \\ 
MG3560n015 & -0.270 &  0.022 &  0.314 &  0.320 \\ 
MG3560n025 & -0.259 &  0.005 &  0.323 &  0.336 \\ 
MG3560p005 & -0.316 & -0.051 &  0.338 &  0.356 \\ 
MG3560p015 & -0.203 & -0.096 &  0.298 &  0.311 \\ 
MG3560p025 & -0.145 & -0.142 &  0.294 &  0.322 \\ 
MG3570n005 & -0.217 &  0.477 &  0.377 &  0.520 \\ 
MG3570n015 & -0.263 &  0.030 &  0.316 &  0.330 \\ 
MG3570n025 & -0.221 &  0.009 &  0.321 &  0.325 \\ 
MG3570p005 & -0.256 &  0.123 &  0.316 &  0.402 \\ 
MG3570p015 & -0.204 & -0.103 &  0.305 &  0.319 \\ 
MG3570p025 & -0.128 & -0.165 &  0.304 &  0.332 \\ 
MG3580n005 & -0.324 &  0.567 &  0.354 &  0.565 \\ 
MG3580n015 & -0.177 &  0.042 &  0.304 &  0.318 \\ 
MG3580n025 & -0.155 &  0.000 &  0.310 &  0.324 \\ 
MG3580p005 & -0.225 &  0.365 &  0.319 &  0.528 \\ 
MG3580p015 & -0.165 & -0.068 &  0.288 &  0.328 \\ 
MG3580p025 & -0.154 & -0.149 &  0.296 &  0.317 \\ 
MG3590n005 & -0.029 &  0.390 &  0.387 &  0.536 \\ 
MG3590n015 & -0.249 &  0.065 &  0.313 &  0.315 \\ 
MG3590n025 & -0.143 &  0.021 &  0.315 &  0.323 \\ 
MG3590p005 & -0.249 &  0.596 &  0.357 &  0.507 \\ 
MG3590p015 & -0.219 &  0.091 &  0.311 &  0.352 \\ 
MG3590p025 & -0.103 & -0.133 &  0.289 &  0.321 \\ 
MG3600n025 & -0.163 &  0.066 &  0.298 &  0.320 \\ 
MG3600p025 & -0.089 & -0.091 &  0.303 &  0.341 \\ 
\enddata
\end{deluxetable}

%% file: confuse_table1.tex
\begin{deluxetable}{ccccc}
\tabletypesize{\scriptsize}
\rotate
\tablecaption{MIPSGAL Confusion Conditions \label{confuseflag}}
\tablewidth{0pt}
\tablehead{\colhead{Region No.} & \colhead{Conditions} & \colhead{No. w/o GLIMPSE} & \colhead{No. w/ GLIMPSE} & \colhead{Ratio}}
\startdata
1 & $d_{NN}$$<$5\arcsec\ OR ($d_{NN}$$<$8\arcsec\ AND $|\Delta$[24]$|$ $<$ 0.2~mag) & 2083 & 7197 & 0.290 \\
2 & 8.0\arcsec$<$$d_{NN}$$<$11.5\arcsec\ AND $\Delta$[24]$>$0.8~mag AND $\Delta$[24]$>$(0.4($d_{NN}$-11.5)+1.8) mag & 3284 & 10767 & 0.305 \\ 
3 & 25\arcsec$<$$d_{NN}$$<$27.5\arcsec\ AND $\Delta$[24]$>$5.4~mag & 355 & 600 & 0.591 \\
Ref. & 13\arcsec$<$$d_{NN}$$<$25\arcsec\ AND $|\Delta$[24]$|$ $<$ 5.4 mag & 35655 & 393842 & 0.091 \\
\enddata
\end{deluxetable}

%% file: cat_table1.tex
\begin{deluxetable}{c}
\tabletypesize{\scriptsize}
\tablecaption{MIPSGAL Catalog Requirements Summary\label{catflg}}
\tablewidth{0pt}
\tablehead{}
\startdata
$\sigma[24] < 0.2$~mag \\
$|FWHM - 6.25$\arcsec$| < 0.5$\arcsec$ (1 + 0.125 \times [24])$ \\
Internal Confusion Flag $= 0$ \\
Edge Flag $= 0$ \\
\enddata
\end{deluxetable}